\documentstyle[aps,epsf]{revtex}

\def\epsfig#1#2#3#4
         {
         \epsfysize=#2 \vbox{ \hglue#3 \epsfbox[#4]{#1} }
         }
\def\epsfigrot#1#2#3#4
         {
         \epsfxsize=#2
         \setbox\rotbox=\hbox to #2{\epsfbox[#4]{#1}}
         \vbox{\hglue#3 \rotl\rotbox}
         }
\newbox\rotbox

\begin{document}

\title{Tunneling in quantum wires II:
A new line of IR fixed points.}
\author{F. Lesage, H. Saleur, P. Simonetti.}
\address{Department of Physics, University of Southern California, 
Los-Angeles, CA 90089-0484.}
\date{\today}
\maketitle

\begin{abstract}
In a previous paper, we showed that 
the problem of tunneling in quantum wires was integrable in the
isotropic case $g_\sigma=2$. In the present work, we continue the exploration
of the general phase diagram by looking for other integrable 
cases. Specifically, we discuss in details the manifold $g_\rho+g_\sigma=2$,
where the associated ``double sine-Gordon'' model is integrable. 
Transport properties are exactly computed. 
Surprisingly, the IR fixed points, while having complete reflection of charge 
and spin currents, do not correspond to two separate leads. Their main
characteristic is that they are approached along irrelevant operators 
of dimension $1+{1\over g_\rho}$ and $1+{1\over g_\sigma}$,
corresponding to transfer of one electron charge but no spin, or one spin $1/2$
but no charge.

\end{abstract}

\section{Introduction.}

While much progress has been made recently in the field of quantum impurity 
problems, it is still an open challenge to fully understand
the phase diagram for  a single
(charge) impurity
in a  one dimensional quantum wire, where the electrons are described
by a Luttinger liquid with charge and spin degrees of freedom.

The problem was studied using the renormalisation group and
perturbation theory by Kane and Fisher \cite{kanefish}
and Furusaki and Nagasoa \cite{furusaki}.  A
complete picture of the flows generated by the RG was
obtained in terms of the ``$g$" factors, $g_\rho$ and
$g_\sigma$, describing the interaction of  bulk charge and spin
degrees of freedom. The existence of tentalizing new fixed points,
which would be partly
transmitting and partly reflecting, was in particular conjectured.
While steps   were taken to identify these fixed points using conformal
field theory \cite{AW}, the determination of their exact nature - 
in particular the computation of the universal associated conductances -
remains open.

In this paper,  we are addressing this impurity problem from the
point of view of integrability.  A simpler problem, 
the case of an impurity in a spinless electron gas, was
mapped on the boundary sine-gordon model \cite{flsbig}, which is a
completely integrable quantum field theory \cite{GZ}.  In our case,
the problem of electrons with spin and impurity, 
it maps after bosonization and the same folding
maneuvers as \cite{flsbig} on a ``double sine-Gordon model'', with two
coupling constants $\beta_1$ and $\beta_2$ (related simply to
$g_\rho$ and $g_\sigma$). Unfortunately, this model does not seem to
be generically integrable. There are however special
integrable manifolds.
In the first part of this work \cite{prospero}
the case $g_\sigma=2$ was solved
by using a slightly different theory than the double sine-Gordon model,
with however the same boundary interaction.  Here a more direct
route is adopted and we investigate the integrability of the
``double sine-Gordon models".
The simplest one is given by $g_\rho+g_\sigma=2$, and can
be considered as a particular case of a new integrable model recently
discovered by Fateev \cite{Fateev}.

This integrable manifold is, a priori, not the most interesting,
since it lies far away from the region where the new IR fixed points
are conjectured to
arise. However, a closer study along the lines of \cite{flsbig} gives rise
to very intriguing results. By a variety of arguments, including the
computation of the non equilibrium tunneling current at vanishing
temperature, we found that  the IR
fixed point does {\bf not} correspond to two separate leads (``open''
fixed point \cite{AW}), as suggested   by a (maybe too naive)
analysis of the classical action  \cite{kanefish}. It is not completely
clear to us what this IR fixed point is. Although it corresponds to
complete reflection (vanishing transmitted current)  for spin and
charge, the operators along which
it is approached seem to correspond to collective excitations that
would
be impossible for separate leads: the leading operators describe
actually
either transfer of one electron charge but no spin, or one spin $1/2$
but no charge!

The paper is organized as follows. Section 2 is a brief summary of
the field theoretic description of the problem and of the folding of
\cite{flsbig}. Section 3 studies the ``double sine-Gordon model'' in
full generality. The integrable manifold is studied using the
technique of non local conserved currents of \cite{BL}, and S and R
matrices are obtained. The IR fixed point is (partially) identified
and shown to differ from the expected one. In section 4, the out of
equilibrium current with an external potential is computed exactly.
Section 5 contains our conclusions. Many additional technical details
are
given in appendices. In appendix A, we explain why the model,
although quantum integrable, is presumably not classically integrable. This is
in sharp contrast with the
case of the ordinary sine-Gordon model, and  suggests that semi
classical arguments as in \cite{kanefish} are maybe not reliable.
Appendix B discusses the problem of quantum  boundary integrability.
In appendix C, a perturbative computation of the S matrix around the
free fermion point is presented, which confirms our non perturbative
identification based on non local currents. Appendix D is a first
order Keldysh computation that fixes the
relation between the bare coupling in our action and the parameter
$T_B$
of the R matrix. Appendix E sketches the computation of the whole two
point current correlator using form-factors, which allows a more
complete identification of the operator content of the IR fixed
point.

\section{The model.}

In order to fix notations and the problem, let us repeat
some of  the definitions in \cite{kanefish}.  
In the one dimensional Luttinger liquid, bosonisation of the
fermionic operators is accomplished via
\begin{equation}
\psi_\mu^\dagger \simeq \sum_{n \ odd} e^{i n[\sqrt{\pi} \Theta_\mu
+k_F x]} e^{i\sqrt{\pi} \phi_\mu(x)},
\end{equation}
with $\mu=\uparrow , \downarrow$.  The fields $\phi_\mu$ and
$\Theta_\mu$ have commutation relations
\begin{equation}
[\phi_\mu(x),\Theta_{\mu'}(x')]=i \delta_{\mu ,\mu'} \theta(x-x')
\end{equation}
from which we can devise two different representations of the
Luttinger liquid
(our conventions follow \cite{kanefish}).
Here, we will work in the so called $\Theta$-representation.
Changing basis to the charge and spin degrees of freedom
\begin{equation}
\Theta_\rho=\Theta_\uparrow+\Theta_\downarrow , \ \
\Theta_\sigma=\Theta_\uparrow-\Theta_\downarrow ,
\end{equation}
we obtain the action
\begin{eqnarray}
S&=&S_\rho+S_\sigma  \\
&=&\int dx dy \frac{1}{2g_\rho} [(\partial_x \Theta_\rho)^2+
(\partial_y \Theta_\rho)^2]+
\frac{1}{2g_\sigma} [(\partial_x\Theta_\sigma)^2+
(\partial_y\Theta_\sigma)^2]. \nonumber
\end{eqnarray}
where $y$ is the imaginary time,  the spin and charge velocities
have been normalised
such that $v_\sigma=v_\rho=1$.  In this convention,
the g-factors have value
$g_\sigma=g_\rho=2$ for a non-interacting system.

The electric and magnetic conductances of this
system follow directly from Kubo's formula
\begin{equation}
\label{old}
G_{\rho / \sigma}=\frac{e^2}{h} g_{\rho / \sigma}.
\end{equation}
As discussed in \cite{res}, this formula is actually not correct
for the charge conductance, 
because of the way the system is coupled to the reservoirs. 
One rather has
\begin{equation}
\label{correct}
G_\rho=2\frac{e^2}{h},
\end{equation}
for any $g_\rho$.  
In this paper, we compute the conductance as  in
\cite{kanefish} to make comparison with the results of this paper more
straightforward.  The correct physical  results follow by  simple
rescaling.

We are interested in
the physical properties of this
model in the presence of a charge impurity
at the origin $x=0$.
Due to this impurity,  the hamiltonian gets an additional piece
\begin{equation}
\delta H=\int dx V(x) (\psi^\dagger_\uparrow \psi_\uparrow+
\psi^\dagger_\downarrow \psi_\downarrow ).
\end{equation}
Here $V(x)$ is a potential which has essentially zero measure outside
$x=0$ and under the bosonisation rules sketched at the beginning,
this leads to the change of action
\begin{equation}
\delta S=\lambda \int dy \cos\sqrt{\pi}\Theta_\rho
\cos\sqrt{\pi}
\Theta_\sigma ,
\label{chagact}
\end{equation}
where $\lambda\simeq V(2k_F)$.
In general, all terms consistent with the
symmetries of the action must be considered at the boundary since
they will be generated by renormalization.   This means that the
most general form of the perturbation is
\begin{equation}
\delta S=\int dy \sum_{n_\rho, n_\sigma\atop n_\rho+n_\sigma \ even}
\frac{\lambda_{n_\rho,n_\sigma}}{4} e^{i\sqrt{\pi}(n_\rho
 \Theta_\rho+n_\sigma
\Theta_\sigma)}
\end{equation}
where the $\lambda$'s are real couplings.  The renormalisation
group equations read at  first order
\begin{equation}
\frac{d\lambda_{n_\rho,n_\sigma}}{dl}=
\left( 1- \frac{n_\rho^2}{4} g_\rho-\frac{n_\sigma^2}{4} g_\sigma
\right) \lambda_{n_\rho,n_\sigma }.
\end{equation}
In this paper, for reasons given later, we will restrict to the manifold
\begin{equation}
g_\sigma+g_\rho=2,
\end{equation}
so that only the perturbation with $n_\rho=n_\sigma=1$ is relevant.
In the following we denote $\lambda_{11}$ by $\lambda$.

In \cite{flsbig}, the equivalent problem for spinless electrons
was exactly solved. The solution required  a folding to transform the
impurity
problem into a boundary problem, and then used recent results
on boundary integrable quantum field theories \cite{GZ} together with
the massless scattering approach. The same folding can easily
be acomplished in the problem with spin. First, it is convenient
to  rescale the fields,
writing the action as
\begin{eqnarray}
\label{lateruse}
S&=&S_\rho+S_\sigma+\lambda \int dy \cos\sqrt{\pi g_\rho}\Theta_\rho
\cos\sqrt{\pi g_\sigma}\Theta_\sigma \\
S_\mu&=&\int dx dy \frac{1}{2} [(\partial_x\Theta_\mu)^2+
(\partial_y \Theta_\mu)^2].
\end{eqnarray}
We then introduce  odd and even fields
\begin{eqnarray}
\Theta^e_\mu&=&\frac{1}{\sqrt{2}} [\Theta_{\mu, L}(x,y )+\Theta_{\mu,
R}
(-x,y)] \\
\Theta^o_\mu&=&\frac{1}{\sqrt{2}} [\Theta_{\mu, L}(x,y )-\Theta_{\mu,
R}
(-x, y )]. \nonumber
\end{eqnarray}
With this, the interaction at $x=0$ involves only the even fields
which have
Neumann boundary conditions.  The odd fields, having Dirichlet
boundary
conditions, completely decouple and do not interact.  In the previous
form, the even field is left moving and we can ``fold" using
\begin{eqnarray}
\theta_{\mu, L}&=& \Theta_\mu^e(x,y), \ x<0 \nonumber \\
\theta_{\mu,R}&=& \Theta_\mu^e(-x,y), \ x<0,
\end{eqnarray}
and now expressing everything in terms of the fields $\theta$ which
is defined on the negative axis, we get the action
\begin{eqnarray}
\label{bdaction}
S&=&S_0+{\cal B} \nonumber \\
&=&\int_{-\infty}^0 dx \int dy \frac{1}{2} \sum_{\mu=\sigma,\rho}
[(\partial_
x\theta_\mu)^2
+(\partial_y\theta_\mu)^2]+\lambda \int dy
\cos\sqrt{\frac{\pi g_\rho}{2}} \theta_\rho (0)
\cos\sqrt{\frac{\pi g_\sigma}{2}}\theta_\sigma (0).
\label{action}
\end{eqnarray}
The next step in \cite{flsbig} was to use the integrability of the
corresponding boundary quantum field theory - the boundary
sine-Gordon model
 (BSG). In the present case with spin, the
problem, involving two fields, is more complicated,
and  very few results are available. We must first investigate in
more details
this ``double sine-Gordon model'' (DSG) and its boundary counterpart
(BDSG).
The following section, together with a series of appendices, are
devoted to
that study. The reader interested in physical properties only
can go directly to the end of next section.

\section{The double sine-Gordon model}

\subsection{Non local currents and integrable manifold}

Although the problem at hand involves an interaction
only at the boundary, it is crucial - like for the ordinary
sine-Gordon model \cite{FSW} -
to understand  first the model with a bulk interaction. We start thus
with the
general action for the double sine-Gordon model, with two fields
$\phi_1, \phi_2$
\begin{eqnarray}
\label{bulkact}
S&=&S_0+S_{bulk} \nonumber \\
&=&S_0+\Lambda
\int dx dy \cos\beta_1\phi_1 \cos\beta_2\phi_2.
\end{eqnarray}
A more general form of this action was first investigated
in \cite{Fateev}, where an integrable manifold was identified, of which
our model is a subset. The following arguments, in particular the use
of affine
quantum group symmetries, are ours, but were strongly influenced by
the results
 of \cite{Fateev}.
The double cosine term (the perturbation)  has dimension
$x={\beta_1^2+\beta_2^2\over 4\pi}$. The free propagators are
$\left<\phi_i(z,\bar{z})
\phi_j(z',\bar{z}')\right>=-{1\over 4\pi}\delta_{ij}\ln |z-z'|^2$,
where we set
$z=y-ix$.

Unfortunately, we do not expect  the model to be integrable
for any choice of
the coupling constants $\beta_1,\beta_2$.  A useful strategy to
identify
quantum integrable varieties is to look for non-local conserved
currents, following Bernard and Leclair \cite{BL}. A current $J(z)$
will be conserved if it is local with
respect to the perturbation, and if
the residue of the simple pole in the short distance
expansion of $J$ and the perturbation is a total derivative. By trial
and error, the simplest possible candidates have the form
\begin{equation}
\label{currents}
J_i^\pm(z)=\partial\varphi_i\ e^{\pm i {4\pi\over
\beta_j}\varphi_j},
\end{equation}
where $\varphi_i(\bar{\varphi})$ are the chiral and antichiral
components of the field,
and in (\ref{currents}), $i\neq j$. One has for instance
(exponentials are implicitely normal ordered)
$$
\partial\varphi_1(z)e^{-i{4\pi\over\beta_2}\varphi_2(z)}
e^{i[\beta_1\varphi_1(w)
+\beta_2\varphi_2(w)]}={1\over z-w}
e^{i[\beta_1\varphi_1(w)+\beta_2\varphi_2(w)-
{4\pi\over\beta_2}\varphi_2(z)]}\left
[-{i\beta_1\over 4\pi (z-w)}+\partial\varphi_1\right]
$$
so the simple pole has residue
$$
\left(-{\beta_1\over
\beta_2}\partial\varphi_2+\partial\varphi_1\right)
e^{i[\beta_1\varphi_1+(\beta_2-
{4\pi\over\beta_2})\varphi_2]}.
$$
This is a total derivative iff $-{\beta_1^2\over
\beta_2}=\beta_2-{4\pi\over \beta_2}$
or
\begin{equation}
\label{cond}
\beta_1^2+\beta_2^2=4\pi.
\end{equation}
In that case, the perturbation has physical  dimension $d=1$
\footnote{In the case
of the ordinary sine-Gordon model,
this dimension corresponds to the free fermion point. The manifold
(\ref{cond})
does not, however, correspond to free fermions except
when $\beta_i^2=2\pi$.}.  Setting
\begin{equation}
\bar{\partial}J^\pm_i=\partial H^\pm_i; \
\partial\bar{J}_i^\pm=
\bar{\partial}
\bar{H}_i^\pm,
\end{equation}
the currents give rise to conserved charges
\begin{eqnarray}
Q^\pm_i\propto \int dz
J^\pm_i+\int d\bar{z}H_i^\pm\nonumber \\
\bar{Q}^\pm_i\propto \int d\bar{z}
\bar{J}^\pm_i+\int dz\bar{H}_i^\pm.
\end{eqnarray}
By counting arguments, following \cite{Zamo}, one expects that the
conservation extends to all orders in perturbation theory \footnote{The fact that 
the dimension of the perturbation is a rational number 
weakens the argument of \cite{Zamo}.} .
In addition, we of course have two conserved topological charges
\footnote{Note
that we divide by $\pi$, not $2\pi$ as in the sine-Gordon model.}
\begin{equation}
T_i={\beta_i\over\pi}\int_{-\infty}^\infty \partial_x
\phi.
\end{equation}
The invariance of the action  constrains $T_1$ and $T_2$ to
have the same parity.

Following the construction of \cite{BL}, one checks that the
non local conserved charges   generate the  quantum
affine algebras $\widehat{sl_{q_i}}(2)$,
with
\begin{equation}
q_i=-\exp\left(-i{2\pi^2\over \beta_i^2}\right).
\end{equation}
Moreover, one also checks that charges of different  type $i\neq j$
commute. For instance, one has
$$
\partial\varphi_1(z)e^{i{4\pi\over\beta_2}\phi_2(z)}
\partial\varphi_2(w)
e^{-i{4\pi\over\beta_1}\varphi_1(w)}=\left[{i\over
\beta_1(z-w)}+\partial
\varphi_1\right]\left[{i\over\beta_2(z-w)}+\partial\varphi_2\right]
e^{i{4\pi\over\beta_2}\varphi_2
(z)}e^{-i{4\pi\over\beta_1}\varphi_1(w)}
$$
so the simple pole has residue
$$
i\left[\left(1-{4\pi\over\beta_2^2}\right)
{\partial\varphi_2\over\beta_1}
+{\partial\varphi_1\over\beta_2}\right]
e^{4i\pi[{\varphi_2\over\beta_2}-
{\varphi_1\over\beta_1}]
}
$$
which is a total derivative on the manifold (\ref{cond}).

Again, while the commutation relations are established to first
order in perturbation theory, simple scaling arguments show that
they hold
to any order. We conclude that, to any order in perturbation, the
theory exhibits the symmetry
$\widehat{sl_{q_1}}(2)\otimes\widehat{sl_{q_2}}(2)$.

Under the condition (\ref{cond}), it is lengthy but straightforward
to show the existence of non trivial local conserved quantities
(this  has  first been shown in \cite{Fateev}).
Besides the stress energy tensor, the next non trivial
one has dimension 4, and follows
from $\bar{\partial}T_4=\partial H_4$, where
\begin{equation}
T_4=a_1^2(1-3a_1^2)(\partial\varphi_1)^4-\left({1\over 4}-{9\over
4}a_1^2
-4a_1^4\right)a_1^2(\partial^2\varphi_1)^2+(1\to
2)+6a_1^2a_2^2(\partial
\varphi_1\partial\varphi_2)^2
\label{tefour}
\end{equation}
and $a_i^2={\beta_i^2\over 4\pi}$.

The existence of this non trivial conserved quantity is
expected to be enough, following
standard arguments \cite{ZZ} to guarantee quantum integrability.
We can  make an easy guess for the particle excitations
and the factorized S matrix. Commutation
with the two quantum affine algebras independently dictates
a tensor product structure. Moreover, since
$T_1,T_2$ must  always have the same parity, excitations
should be massive particles carrying a {\bf pair} of quantum numbers
$T_1,T_2$:
ie they are simultaneously  solitons of the field $\phi_1$ and
$\phi_2$. One then obtains as the simplest choice
\begin{equation}
\label{SSmatrix}
S=-S_{\widehat{\beta_1}}\otimes S_{\widehat{\beta_2}},
\end{equation}
where $S_\beta$ is the S-matrix of solitons for the ordinary
sine-Gordon model
at coupling $\beta$,
\begin{equation}
{\widehat{\beta_i}^2\over 8\pi}={\beta_i^2\over
\beta_i^2+2\pi},
\end{equation}
and we used the results of \cite{BL}.  This scattering matrix
was first found in \cite{Fateev} using a different argument.
It is important to stress that we do not have independently solitons
of the field $\phi_1$ and of the field $\phi_2$, but composite
objects. Also, the coupling in the S-matrices
is renormalized non trivially. This is largely the consequence of the
$\partial\varphi$ term appearing in the conserved currents, a feature
which is not there
in the standard sine-Gordon model, but is necessary here because the
interaction is
the product of two cosines: although the S-matrix is factorized,
the theory is thus highly interacting. It is interesting to
discuss
the action and the factorized S matrix (\ref{SSmatrix}) from the
point of view
of a fermion theory: see appendix C for more details.

Some special points can be identified on the integrable
manifold, which can be solved independently. For instance
the case  $\beta_1=\beta_2=\sqrt{2\pi}$ can be recast, by forming
linear combinations
of the fields, into a problem of two decoupled sine-Gordon theories
at their free fermion point, in agreement with (\ref{SSmatrix}).
The case $\beta_1=0$ is also trivially solvable, and equivalent to
one free boson and one sine-Gordon model at the free fermion point.
This is a very different description from the one provided by
(\ref{SSmatrix}), which
involves in particular a sine-Gordon S matrix at the point
$\widehat{\beta_2}^2={16\pi\over 3}$,
ie the $N=2$ supersymmetric point. Nevertheless, we will check that,
when physical quantities are computed, they have identical
expressions in both descriptions.

{}From the  relation
\begin{equation}
\left[T_i,e^{i(\alpha\varphi_j+\bar{\alpha}\bar{\varphi}_j)}
\right]=
\delta_{ij}{\beta_i\over 2\pi}(\alpha-\bar{\alpha})
 e^{i(\alpha\varphi_j+\bar{\alpha}\bar{\varphi}_j)},
\end{equation}
we find the soliton operators (up to polynomials in derivatives of
$\phi_1,\phi_
2$)
\begin{eqnarray}
\Psi_i^\pm=&\exp\left[\pm
i{2\pi\over\beta_i}\varphi_i\right]\nonumber \\
\bar{\Psi}_i^\pm=&\exp\left[\mp
i{2\pi\over\beta_i}\bar{\varphi}_i\right],
\end{eqnarray}
which have topological charge $T_i=\pm 1$.
The local current $J_i^\pm$ is
a raising (lowering) operator for particles of type $i$: it
annihilates
solitons (antisolitons)  and transforms antisolitons (solitons)
into solitons (antisolitons). $J_i^\pm$ conserves $T_j$, $i\neq j$.

\subsection{Boundary integrability and  new IR fixed points}

The main  step in \cite{flsbig} was  to use the integrability
of the boundary sine-Gordon model.
Here, one can easily prove that the model (\ref{bulkact})  
defined on the half
line $x\in [-\infty,0]$ with a boundary interaction at $x=0$
\begin{equation}
\label{bdract}
{\cal B}=\lambda\int dy \cos{\beta_1\over 2}\phi_1(y)
\cos{\beta_2\over 2}\phi_2(y),
\end{equation}
is  integrable for any choice $\lambda,\Lambda$. This model we call
the double boundary sine-Gordon model (DBSG). In the factorized
scattering picture,
the boundary interaction is described by a  reflection matrix
solution of the boundary Yang Baxter equation. The
factorized structure of the bulk S-matrix leads to the immediate
conjecture for this  reflection matrix
\begin{equation}
R=-R_{\widehat{\beta_1}}\otimes R_{\widehat{\beta_2}},
\end{equation}
where $R$ is the reflection matrix of the boundary sine-Gordon
model given explicitely in \cite{GZ}.

Of course the original Kane-Fisher impurity
problem is massless in the bulk.
As in the spinless case,
we describe this problem starting from the massive theory with
a bulk coupling (\ref{bulkact}) by using the quasiparticles basis,
and letting $\Lambda \to 0$. In that limit, we get  right and left
moving
massless solitons with dispersion relation $e=\pm p=me^\theta$,
with $\theta$ the rapidity. Movers of opposite
types interact via a constant S-matrix, while movers of the same type
still interact
with (\ref{SSmatrix}). In addition, R movers bounce back as L movers
at the boundary.  We recall in particular the physical amplitudes
\begin{equation}
\label{ampl}
\left|(R_{\widehat{\beta_i}})^+_+\right|^2=
{1\over 1+e^{-\frac{4\pi}{\beta_i^2}(\theta-\ln T_B)}},\ \ \ 
\left|(R_{\widehat{\beta_i}})^+_-\right|^2
=1-\left|(R_{\widehat{\beta_i}})^+_+
\right|^2.
\end{equation}
Here, $T_B$ is an energy scale related with the coupling $\lambda$ by
$T_B\propto \lambda^{2}$,
where we used (\ref{cond}).

Already at this point, the identification of these amplitudes allows
a rough analysis of the infrared fixed point.
If we expand the modulus square of the
R-matrices in terms of the  inverse coupling
we find that $|R_{\widehat{\beta_i}}|^2$ expands in
powers of $T_B^{-4\pi/\beta_i^2}\propto \lambda^{-8\pi/\beta_i^2}$. On
the other hand,  by the Fermi Golden rule, these  are proportional
to the square of the matrix elements
of the  operators perturbing the IR fixed point. We thus see that
these operators must have dimension
\begin{equation}
\label{dimi}
d_i={2\pi\over\beta_i^2}+1.
\end{equation}
The same result is obtained  when analyzing physical properties (see
below).

This conclusion is, a priori,
very surprising. Indeed, the standard  analysis  \cite{AW}
of the IR limit based on
the classical
action (\ref{bdract}) would suggest that, at large coupling, the
fields
$\phi_i$ are located at the bottom of the boundary potentials,
leading to
Dirichlet boundary conditions with
\begin{equation}
\phi_i(x=0)={2\pi\over \beta_i} n_i,\ n_i+n_j=0\hbox{ mod } 2.
\label{diri}
\end{equation}
This fixed point we refer to in the following as the standard IR
fixed point. With (\ref{diri}), it is easy to classify the allowed
operators
describing the approach to the  standard IR fixed point. One simply
computes the partition function of the model with these boundary
conditions at $x=0,x=L$, and matches with the formula of conformal
invariant field theories, $Z=Tr w^{L_0-c/24}$. Here, one finds
\begin{equation}
Z={1\over \eta(w)}\sum_{n_1,n_2}' w^{-2\pi
\left({n_1^2\over\beta_1^2}+{n_2^2\over\beta_2^2}\right)},
\end{equation}
where the primed sum indicates that $n_1,n_2$ have the same parity,
$w=e^{-\pi/L
T}$.
The dimensions of the allowed operators,
up to integers, are thus of the form
$$
d=2\pi\left({n_1^2\over\beta_1^2}+{n_2^2\over\beta_2^2}\right).
$$
In particular, the lowest dimension operators have dimension
${8\pi\over \beta_i^2}, {2\pi\over \beta_1^2}+{2\pi\over \beta_2^2}$.
Remarkably,
the ones with
$x={2\pi\over \beta_i^2}+1$ are
not there!

There are several explanations for this discrepancy. The first, of
course, is the possibility that our S and R  matrices  have not been
correctly identified.
The previous  arguments about affine quantum group symmetries
are however usually considered to be quite strong. They can be made
even stronger 
at some special points with $N=2$ supersymmetry (see below).
Moreover, physical results for all  limiting cases where the
solutions is known by other means will
be reproduced correctly (see  also below). Finally, the S and R
matrices
also agree with perturbation theory
 around the free fermion point $\beta_i^2=2\pi$ (see appendix C).
If we do believe these S and R matrices, another
possible explanation is that the perturbation around the IR fixed
point has a vanishing radius of convergence, so the expansion
of (\ref{ampl}) (or other physical
quantity) in powers of $T_B$ is the result of a resummation, and
cannot be used to  naively identify the dimension of the perturbing
operator. Such phenomenon has been observed in other integrable
 models \cite{saus}.

We suspect however that this discrepancy arises, more simply,
because the IR fixed point obtained by naive analysis of the action is
not the  correct one. To justify this, we recall first that integrable
flows are usually found to be integrable considered  either as perturbations
of the UV or of the IR 
fixed point, and with the same set of conserved quantities. For instance,
 in the boundary sine-Gordon model perturbed by
$\cos\beta\phi$, the local conserved quantities like $T_4$ commute not only 
with the perturbation, but also with its ``dual'' $\cos {8\pi\over \beta}\phi$. 
As a result, the flow near the IR fixed point, which is approached along this 
dual field plus local integrals of motion $T_{2n}$, is integrable too. 
In the present 
case, we checked that our $T_4$ quantity does commute, besides $\cos\beta_1\phi_1
\cos\beta_2\phi_2$,  with
the currents $J_i^\pm$, but {\bf not} with $\cos{8\pi\over \beta_i}\phi_i$,
nor with $\cos {4\pi\over \beta_1}\phi_1\cos{4\pi\over \beta_2}\phi_2$.
As a result, it seems impossible to flow to the disconnected fixed point,
since its approach would not be integrable. On the other hand,
the  IR fixed point described above was found to be
approached along  operators of the same dimension as $J_i^\pm$. Very presumably,
it is approached along these operators indeed (plus the $T_{2n}$), providing
an integrable flow. 

As another piece of justification,  it is useful to consider a generalization of
(\ref{bulkact})  given by the action
\begin{equation}
S={1\over 2}\int dxdy \left[(\partial_\mu \phi_1)^2+
(\partial_\mu \phi_2)^2+(\partial_\mu \phi)^2\right] +
{\Lambda\over 2}\int dx dy \left[
e^{\beta\phi}\cos(\beta_1\phi_1+\beta_2\phi_2)+e^{-\beta\phi}
\cos(\beta_1\phi_1-\beta_2\phi_2)\right]
\label{newact}
\end{equation}
involving thus one more bosonic field. Non local conserved currents
generalizing  (\ref{currents}) can be built
\begin{equation}
J_i^\pm (z)=\left(i\beta_i\partial\varphi_i\pm \beta\partial\varphi
\right)
e^{\pm i{4\pi\over\beta_j}\varphi_j}
\end{equation}
provided the condition
\begin{equation}
\beta_1^2+\beta_2^2-\beta^2=4\pi,
\label{newcond}
\end{equation}
holds. By the same arguments,
the model is integrable. An  integrable boundary
action follows
\begin{equation}
{\cal B}={\lambda\over 2}\int dy \left[
e^{\beta\phi(0)/2}\cos\left({\beta_1
\phi_1+\beta_2\phi_2\over 2}(y)\right)
+e^{-\beta\phi(0)/2}\cos\left({\beta_1
\phi_1+\beta_2\phi_2\over 2}(y)\right)\right].
\label{newbdact}
\end{equation}
It is possible to compute the algebra generated by the
conserved charges and it is found to be the same 
as in the previous model.
Thus, the S and R matrices for this model are essentially the same
as the ones for the  previous,  simpler model with $\beta=0$,
with however the  the new condition  (\ref{newcond}).
The  IR fixed point deduced from the classical action
is the ``standard'' one like for $\beta=0$
since the field $\phi$ cannot be compactified.

This generalization is useful because at the
value $\beta_1^2=4\pi$, the model (\ref{newact})
coincides with the bosonized form of the $N=2$ supersymmetric 
sine-Gordon model,
as can easily be inferred from the fact that the $J_1$ currents have
dimension
$3/2$: $J_1^\pm\equiv G^\pm$. The bulk perturbation is unitary and
preserves supersymmetry \cite{KU}. The same is true for the boundary
perturbation. This
supersymmetry is a particular case of the quantum affine symmetry
generated by the
currents $J_i^\pm$ . The preservation of supersymmetry is a more
firmly based
property than the preservation of quantum affine symmetries however,
because the
generators are local. We now use the conservation
of supersymmetry to justify the identification of the IR
fixed point provided by the S and R matrices.  Let us  first
recall what happens for the supersymmetric point
of the ordinary  boundary sine-Gordon model first \cite{W}. Then,
$\beta^2={16\pi\over 3}$. The
perturbing operator $\cos\beta\phi$ in the UV has
boundary dimension $x=2/3={1\over 2}+{1\over 6}$
corresponding to the superpartner of the chiral primary field $X$,
and preserves
 supersymmetry.
The IR fixed point is, in that case, correctly reproduced by
the standard  analysis of the classical action.
The Dirichlet boundary conditions analogous to (\ref{diri})
give rise to the partition function
\begin{equation}
Z={1\over\eta(w)}\sum_n w^{{3n^2\over 2}}=\chi^{NS}_{00},
\end{equation}
which is the Neveu Schwartz character of the identity (that one
is in the NS sector follows immediately from the fact that boundary
conditions are the same on both ends of the cylinder).  The IR fixed
point
is approached along the operator $\cos{8\pi\over\beta}\phi$
of dimension $x=3/2$, ie along the operators
$G^\pm$. This is very natural from a Landau Ginzburg point of view:
near
the minima of the potential, the effective potential is $W=X^2$ with
no left-over  chiral primary field in the spectrum beside the
identity.
The approach has
thus
to be along a descendent of the identity, with $G^\pm$ as
the natural candidates.
It is not totally obvious a priori that a perturbation along
 $G^\pm$  preserves supersymmetry;  while $G^+$ anticommutes with
$G^+$, the
anticommutator of $G^+$ and $G^-$ is non zero. However, the
$G^\pm$ are fermionic operators; to have a bosonic IR
action, they must thus be coupled to some  fermionic  boundary
degrees
of freedom, and these  allow the restoration of supersymmetry
\cite{W}.

We can now get back to the  model (\ref{newbdact}) when
$\beta_1^2=4\pi$, ie
with unbroken $N=2$
supersymmetry. It is easy to check that the standard IR fixed point
(\ref{diri})
cannot be approached while preserving supersymmetry. On the other
hand,
the leading operators describing the approach to our new IR
fixed point have dimensions $d=3/2$ and $d=1+{2\pi\over\beta_2^2}$.
The first
one corresponds to $J_1^\pm=G^\pm$, and, like for the ordinary
sine-Gordon model,
it does preserve supersymmetry after addition of  fermionic boundary
degrees of freedom. The second one corresponds to $J_2^\pm$. Because
the two affine quantum algebras in the problem commute, it too does
preserve supersymmetry.
Hence, our fixed point is perfectly compatible with unbroken
supersymmetry,
while the standard one is not.

Away from the supersymmetric points, a similar argument can be made based
on the conservation of affine quantum symmetries instead
\footnote{This is a little tricky to do in
practice, because of non
locality problems for general values of $\beta_i$. It is however
possible for
special values of $\beta_i$ \cite{LV}.}. It leads right away
to the claim that the IR fixed point should be approached along
the non local conserved currents: $J^\pm\propto e^{\pm 8i\pi\phi/\beta}$
in the ordinary sine-Gordon case, as is well known, and 
along $J_i^\pm$, of dimensions $d_i=1+{2\pi\over\beta_i^2}$, in the 
double sine-Gordon case, 
in agreement with the conjectured S and R matrices .

Based on the analysis of physical properties (see below
and appendix E),
we conjecture that the   operator content of the IR fixed point is
a {\bf  subset} of the spectrum encoded in
\begin{equation}
Z={1\over\eta^2(w)}\sum_{n_1}w^{-2\pi{n_1^2\over\beta_1^2}}
\sum_{n_2}w^{-2\pi{n_2^2\over\beta_2^2}}
\label{conji}
\end{equation}
Note that the partition function (\ref{conji}) would correspond to
a model with Dirichlet boundary conditions of the type (\ref{diri})
but with no coupling between $n_1$ and $n_2$. A possible conjecture
for the ratio of boundary entropies in the UV and IR is
\begin{equation}
{g_{IR}\over g_{UV}}=\left({\beta_1^2\over\beta_1^2+2\pi}
{\beta_2^2\over\beta_2^2+2\pi}\right)^{1/2}
\end{equation}
while the same ratio for the standard fixed point would read
$\left({g_{IR}\over g_{UV}}\right)_{standard}=2\left({\beta_1^2\over 8\pi}
{\beta_2^2\over 8\pi}\right)^{1/2}$. If this result is
correct, our  new fixed point is less stable than the open (separate
leads) one: therefore, it might well be that it is reached only with
the UV action  (\ref{chagact})
on the integrable manifold $g_\rho+g_\sigma=2$.

The question of course remains:  why is the  identification of the IR
fixed point misleading in our case, while it was not for the boundary
sine-Gordon model? We have no satisfactory answer to that question
at the present time. But there is an interesting indication: in sharp
distinction with
the ordinary sine-Gordon model, the double sine-Gordon model
(\ref{bulkact}) seems to not be integrable classically, 
as discussed in the appendix A.
Integrability in the present case
is thus a  truly  quantum property. Since integrability is also
responsible (via the analysis of non local conserved currents 
and the resulting S and R
matrices) of the unexpected  nature of the IR
fixed point, it is maybe not so surprising  that this IR fixed point
cannot be obtained from a classical analysis of the action,
and thus differs from the standard one.

\subsection{Summary}

We can now get back to the physical problem.
A comparison of (\ref{bdaction}) with the results of this
section leads to the identification of the integrable manifold
in terms of the $g$-factors: $g_\rho+g_\sigma=2$.
The postulated exact $S$-matrix has the form of a tensor
product of spin and charge degrees of freedom
\begin{equation}
S=-S_\rho\otimes S_\sigma ,
\end{equation}
with $S_{\rho,\sigma}$ the $S$-matrices of a sine-Gordon model with
parameters ${\beta^2\over 8\pi}={g\over g+1}$. Similarly the
boundary reflection matrix is given by
\begin{equation}
R=-R_\rho\otimes R_\sigma .
\end{equation}
The elementary excitations of the folded theory are
quasiparticles carrying a charge number $
Q_\rho=\pm 1$ (in units where $e=1$)
and a spin number $Q_\sigma=\pm 1$ 
(in units where a physical electron has
$S^z=\pm \hbar$). Here, we have set
\begin{eqnarray}
Q_\rho={1\over \pi}{1\over \sqrt{\pi g_\rho}}\int dx \ \partial_x
\theta_\rho\nonumber\\
Q_\sigma={1\over \pi}{1\over \sqrt{\pi g_\sigma}}\int dx \ \partial_x
\theta_\sigma
\end{eqnarray}

Having  a complete description of the kinematics of the model, we can
now compute physical properties. We will restrict to the $T=0$ case
here.
Extensions to finite $T$ are in principle possible, but quite
technical.

\section{DC conductance at $T=0$}

We will exactly follow the same line of argument as in \cite{flsbig}.
The quasiparticles
are considered as a gas interacting through factorized scattering.
Standard
thermodynamics arguments can be applied, in the presence of  electric
and magnetic chemical potentials, to compute the densities and the
filling
fractions of these quasiparticles. These are expressed via the
thermodynamic
Bethe ansatz. Having these densities,  a rate equation can then
be written to compute the contribution  to the conductances 
of the current backscattered
by the impurity.

\subsection{General computation.}

We only consider the case of an external voltage,
so the quasiparticles have chemical potential $e^{\pm V/2T}$,
$T$ the temperature.
At $T=0$, a Fermi sea is formed,
consisting of positively charged quasi-particles,
with either spin up or down. These quasiparticles carry charge labels
$(1,1)$ and $(1,-1)$. Due to the factorized form of the S matrix,
the scattering of these quasiparticles is factorized into the
scattering
of the first and second labels. The first labels
scatter diagonally with the element $\left(S_\rho\right)^{++}_{++}$.
The second labels in general scatter non diagonally.  They would
scatter
diagonally if $g_\sigma$ was of the form $1/integer$,
but then bound states (carrying a non vanishing electrical charge)
would have to be taken into account. It is in fact simpler
to analyze the more general case corresponding
to model (\ref{newact}), taking $g_\rho=p_1$, $g_\sigma=p_2$,
$p_i$
integers.
The final results can then be analytically continued to the region of
interest (\ref{cond})
with $\beta=0$ (a computation right at $\beta=0$ is also possible,
although more complicated. We will describe this elsewhere).
 The scattering of the magnetic quantum numbers is
then
non diagonal. The TBA is slightly intricate to write: one
needs to introduce pseudoparticles  to diagonalize this scattering.
The final result
fortunately is very much like in the ordinary sine-Gordon model
\cite{FI,KR}, and is conveniently encoded in the TBA diagram
\vskip 0.4cm

\noindent
\bigskip
\noindent
\centerline{
\hbox{\rlap{\raise28pt\hbox{$-p_1+1\hskip.3cm\bigcirc\hskip
4.55cm\bigcirc\hskip .3cm p_2-1$}}
\rlap{\lower27pt\hbox{$-p_1\hskip.8cm\bigcirc\hskip
4.45cm\bigcirc\hskip.3cm p_2$}}
\rlap{\raise14pt\hbox{$\hskip1.55cm\Big\backslash\hskip4.35cm\Big/$
}}
\rlap{\lower14pt\hbox{$\hskip1.5cm\Big/\hskip4.3cm\Big\backslash$
}}
$\hskip1.5cm${\raise1pt\hbox{$\bigcirc$}}------$\bigcirc$-- --
--$\bigotimes$-
--
--$\bigcirc$------$\bigcirc$ }}
\bigskip

The TBA equations are
\begin{eqnarray}
\epsilon_0={1\over 2}e^\theta+{k\over
2\pi}*\left[\epsilon_1^-+\epsilon_{-1}^-\right]\nonumber\\
\epsilon_i={k\over
2\pi}*\left[\epsilon_{i-1}^-+\epsilon_{i+1}^-\right]\nonumber\\
\epsilon_{-p_1}^-=0, \ \epsilon_{-p_1+1}=-p_1{V\over 2}+
{k\over 2\pi}*\epsilon_{-p_1+2}^-
\label{tbaa}
\end{eqnarray}
In these equations, the star indicates a convolution
of the functions with the kernel 
$k(\theta)={1\over 2\cosh \theta}$, and $\epsilon^-$
designates
the negative part of $\epsilon$ \cite{KR}. It is a simple
exercise in Fourier transform to extract from (\ref{tbaa}) an
equation
satisfied by $\epsilon_0$ itself. Setting
$\epsilon\equiv-\epsilon_0^-$,
one has
\begin{equation}
\epsilon(\theta)-\int_{-\infty}^A
\Phi(\theta-\theta')\epsilon(\theta') d\theta'=
{V\over 2}-{1\over 2}e^\theta
\end{equation}
Here, the kernel $\Phi$ has fourier transform
\begin{equation}
\tilde{\Phi}=\tilde{k}^2\left(K_{-p_1+2}+K_{p_2-2}\right)
\end{equation}
where $\tilde{k}(x)={1\over 2\cosh x}$, the kernel $K_{-p_1+2}$
arises from the left part of the diagram, and coincides with the
derivative of the logarithm of $\left(S_\rho\right)^{++}_{++}$,
and the kernel $K_{p_2-2}$ arises from the right part of the diagram
\begin{equation}
K_{-p_1+2}=2\cosh x{\sinh(p_1-1)x\over \sinh p_1x},\ \
K_{p_2-2}=2\cosh x
{\cosh(p_2-2)x\over \cosh(p_2-1)x}
\end{equation}
leading to
\begin{equation}
\tilde{\Phi}=1-{\cosh (p_1+p_2-1)x\over 2\sinh p_1x\cosh (p_2-1)x}.
\end{equation}
We recall that $\epsilon$ describes the excitation energy
of new quasiparticles which do not have a well defined magnetic
quantum number - this is because the scattering of magnetic numbers
is
non diagonal. The total density of these quasiparticles
is obeys
\begin{equation}
\rho(\theta)-\int_{-\infty}^A \rho (\theta ') \Phi(\theta-\theta ')
d\theta '=
\frac{1}{2 h} e^\theta ,
\end{equation}
where we have reinstated Planck's constant. This density
follows simply from the excitation energy as $\rho=-{1\over
h}{d\epsilon\over d\theta}$.

In the previous equations, $A$ is the Fermi rapidity defined by
$\epsilon(A)=0$;
it is the edge of the Fermi sea. $A$ can be determined, and the
foregoing equations
solved, by using a standard Wiener Hopf analysis.
If we write $K(\theta)=\delta(\theta)-\Phi(\theta)$, we have
\begin{equation}
\tilde{K}(\omega)=\frac{1}{N(\omega ) N(-\omega )},
\end{equation}
with
\begin{equation}
N(\omega)=\sqrt{\frac{2\pi}{g_\rho}} \frac{ \Gamma(i\omega /2)
e^{i\omega \Lambda}}{\Gamma(i\omega g_\rho/2)
\Gamma(1/2+i\omega (g_\sigma-1)/2)},
\end{equation}
a function analytic in the lower half plane when we
choose $\Lambda=1/2 \log 2+(g_\sigma-1)/2 \log (g_\sigma-1)/2
+g_\rho/2 \log g_\rho/2$.
Here the conventions for the Fourier transforms are
\begin{equation}
g(\theta )=\int_{-\infty}^\infty \frac{d\omega}{2\pi}
e^{-i \omega \theta} \tilde{g}(\omega), \ \
\tilde{g}(\omega )=\int_{-\infty}^\infty d\theta
e^{i \omega \theta} g(\theta).
\end{equation}
Then following  exactly the same steps as in \cite{flsbig}
we find the solution for the fourier transform
of the density
\begin{equation}
\tilde{\rho}(\omega )=\frac{1}{2 h} \frac{e^{(1+i\omega
)A}}{(1+i\omega)}
N(-i) N(\omega ) .
\end{equation}
The excitation energy of the particles follows
\begin{equation}
\tilde{\epsilon}(\omega) e^{-i\omega A}
=-i \frac{q V}{2} \frac{N(0)N(\omega)}{\omega}
+ \frac{i e^A}{2} \frac{N(-i) N(\omega)}{\omega-i}.
\end{equation}
The condition $\epsilon(A)=0$ is equivalent to
\begin{equation}
\lim_{\omega\rightarrow \infty} \omega \tilde{\epsilon }(i\omega)
 e^{-i\omega A}=0,
\end{equation}
which in turns gives the explicit value of the Fermi rapidity, $A$
\begin{equation}
e^A=e V \frac{N(0)}{N(-i)}.
\end{equation}

In the presence of the impurity, the backscattered current
is computed by using a Landauer B\"uttiker approach \cite{LB}. The
only difference with the computation done in \cite{flsbig}
is that the quasiparticles have a magnetic degree of freedom.
However,
we are only interested here in the properties of the electric
charge, so we simply sum over all possible channels of reflection
for the magnetic quantum number.
This leads to the final expression for the current
\begin{equation}
I(V,T_B)=e \int_{-\infty}^A d\theta \rho(\theta) \
\vert\left( R_\rho\right)^+_+(\theta-\theta_B)\vert^2.
\label{main}
\end{equation}
This expression was derived under the assumption $g_\rho,g_\sigma$
integers.
We expect  however that it holds for any values of these constants,
in particular for
the values on the manifold $g_\rho+g_\sigma=2$. Indeed, physical
quantities in integrable models have (so far) always had a smooth
behaviour in terms of the
coupling constants, despite the very different possible
Bethe ans\"atze that depend on their arithmetic properties.
Furthermore, Fateev \cite{Fateev}
has computed some thermodynamic properties  for the more
 general model
(\ref{newact}). One can check on his computations, by comparison
with perturbative results or sigma models approximations, that
the results are indeed analytical in terms of the coupling constants.

Let us now analyze (\ref{main}). In the UV limit,
the matrices $R_+^+$ become unity and
we find the result
\begin{equation}
I(V,T_B=0)=\frac{ e^2 V}{h} g_\rho ,
\end{equation}
as expected \footnote{To get the correct version, it suffice to
rescale $V$ as mentionned at the beginning}.
The full result at finite $T_B$ is given by a double expansion,
one for each reflection matrix.  The modulus squared of the
previous reflection matrices is given by
\begin{equation}
\vert \left(R_\rho\right)^+_+(\theta)\vert^2={1\over 1+
\exp (-\frac{2}{g_\rho} ) \theta }, \ \
\vert \left(R_\sigma\right)^+_+(\theta)\vert^2={1\over 1+
\exp (-\frac{2}{g_\sigma} ) \theta }
\end{equation}
For large voltage, we just expand the previous expression and
evaluate the current at each order by suming over residues,
we obtain the following strong coupling expansion
\begin{equation}
\label{strong}
I(V,T_B)=\frac{e^2 V}{2h} \sum_{n=1}^\infty I_{2n}
\left(\frac{eV}{T_B}\right)^{2n/g_\rho}
\end{equation}
with
\begin{equation}
I_{2n}= (-1)^{n+1}N(0) \frac{N(-2in/g_\rho )}{(1+2n/g_\rho)}
\left(\frac{N(0)}{N(-i)}\right)^{2n/g_\rho}.
\label{iiis}
\end{equation}
Using a similar procedure and treating the case in which $T_B<V$,
we can also expand the reflection matrices by splitting the
integral over the rapidity, $\theta$ in two pieces.  After a
little algebra, we find the weak coupling expansion
\begin{equation}
I(V,T_B)=\frac{e^2 V}{2h} N(0) PP\int \frac{d\omega}{2\pi}
\frac{N(\omega)}{(1+i\omega)} \left\{
\sum_{n=0}^\infty (-1)^{n+1} \frac{b^n}{2n/g_\rho+i\omega}-
\frac{i\pi g_\rho}{2} \frac{b^{-i\omega g_\rho/2}}{
\sinh (\pi \omega g_\rho/2)}
\right\}
\end{equation}
with $b=(T_B e^{-A})^{2/g_\rho}$.
We integrate the first sum by closing the contour in the lower
half plane and the second term by closing in the upper half plane.
In the second term, the zeroes of $N(\omega)$ cancels the poles
of $\sinh(\pi \omega g_\rho/2)$ and only the poles
$\omega=2ni$ with $n$ integer, and $\omega=i$ contributes.
We get
\begin{equation}
I(V,T_B)=\frac{e^2 V}{h} g_\rho+
\sum_{n=1}^\infty  U_{2n} \left( \frac{T_B}{eV}\right)^{2n}
+U_1 T_B,
\label{iis}
\end{equation}
with the coefficients
\begin{eqnarray}
U_{2n}&=& \frac{e^2 V}{2h} \sqrt{2\pi\over g_\rho} \pi g_\rho
\frac{(-1)^n}{n!} e^{-2n \Lambda}
{N(0) (\frac{N(-i)}{N(0)})^{2n}\over \Gamma[-n g_\rho]
\Gamma[1/2-n(g_\sigma-1)] (2n-1) \sin(\pi n g_\rho)}
\nonumber \\
U_1&=&-\frac{e\pi g_\rho}{2h} \sin (\frac{\pi}{2}g_\rho) .
\end{eqnarray}
Let us write the first few orders more explicitely
\begin{equation}
I(V,T_B)=\frac{e^2 V}{h}g_\rho\left\{ 1-\frac{\pi}{2}
\sin(\frac{\pi g_\rho}{2}) \frac{T_B}{e V}+
\frac{\pi^{5/2}}{g_\rho^2}\frac{\Gamma(g_\rho)}{\Gamma(1/2-g_\rho)}
\frac{1}{\Gamma(g_\rho/2)^2 \Gamma(g_\sigma/2)^2}
\left(\frac{T_B}{eV}\right)^2
+\cdots \right\}
\end{equation}
The spin conductance in a magnetic field follows from the previous
analysis
simply by the interchange of $g_\sigma$ and $g_\rho$.
In order to compare with perturbation theory and complete
the picture, we need to establish a
relation between the boundary scale $T_B$ and the perturbative
coupling $\lambda$.  This is done in appendix D and we find
\begin{equation}
\label{gaprel}
T_B=\frac{\pi}{2\sin(\frac{\pi}{2}g_\rho)} \lambda^2.
\end{equation}

\subsection{Special Cases.}

When $g_\sigma=g_\rho=1$ the sum (\ref{strong})
can be done analytically, in that case we have that
\begin{equation}
I_{2n}= \frac{(-1)^{n+1}}{2n+1}
\end{equation}
and we find the result 
\begin{equation}
\label{spe1}
I(V,T_B)=\frac{ e^2 V}{h}-\frac{ e T_B}{h} \arctan \frac{eV}{T_B}.
\end{equation}
This result can be found by other means since by writing the
action, not in the charge and spin fields basis but rather
in the spin up and down basis, the action decouples into two
Luttinger actions which do not interact with each other.
Moreover each is at the free fermion, or Toulouse, point and
the resulting conductance is well known to be of the form
given by (\ref{spe1}). The two expressions can be precisely matched
using (\ref{gaprel}) together with the results of
\cite{flsbig}\footnote{Most unfortunately, there are two misprints in
the latter reference. In eq. (6.20), the left hand side
should have a $2\lambda A_1$ instead of $\lambda_1$. The next
equation should
have a term $eV/2T_B$ instead of $eV/T_B$ in brackets.}.

Another special point  is $g_\rho=2$ and
$g_\sigma=0$.
The spin field then totally decouples, while
the dynamics of the charge field is described by the
ordinary boundary sine-Gordon model at the free fermion point. One
thus
expects
a conductance of the form (\ref{spe1}) (with however an overall
factor of $2$, and a  renormalized $T_B$). On the other hand, the  S
matrix which we claim describes the double sine-Gordon model
is not at the free fermion point in the charge sector, but rather at
the $N=2$ supersymmetric point! Nevertheless, special cancellations
then occur to reproduce the correct physical quantities. Indeed,
using the expression (\ref{iiis}), we have at that particular point
\begin{equation}
I_{2n}=
(-1)^{n+1}\frac{ N(-in)}{1+n} \left( \frac{2}{N(-i)}\right)^n.
\label{tenti}
\end{equation}
In the limit $g_\rho\rightarrow 2$,
this vanishes for $n$ odd. For $n$ even, one has
\begin{equation}
I_{4n}\approx 2 {(-)^{n+1}\over 2n+1}
\left({1\over \sin \pi g_\rho/2}\right)^{2n}
\end{equation}
This reproduces the correct result using (\ref{gaprel}).

A last special point is $g_\rho=0,g_\sigma=2$. There, one expects
the spin field to essentially decouple as the fluctuations of the
charge field become very large. Explicit computation using
(\ref{iis})
gives rise to
\begin{equation}
{I\over g_\rho}\approx {e^2V\over h}\left(1+\sum_{n=1}^\infty
(-)^n {\sqrt{\pi}\over n! \Gamma({1\over 2}-n)({1\over
2}-n)}\left({\pi T_B\over 2eV}\right)^{2n}\right)
\end{equation}
This coincides indeed with the classical limit of the spinless
problem \cite{flsbig}.

\section{Conclusion}

The main result of our study is the existence of a new kind of IR
fixed point in this impurity problem. This fixed point has a
vanishing
transmitted current but it  does not correspond to the open (separate
leads) fixed point as demonstrated by the analysis of the operator
content. Of course, since the transmitted current vanishes, it does
not correspond to any of the other fixed points (eg current
transmitting and spin reflecting, or partially transmitting and
partially reflecting)  considered so far \cite{kanefish,AW}. The
main property of our fixed point is that it is approached along
operators which
describe the transfer of one electron charge but no spin, or one spin
$1/2$ but no charge. This  is clearly seen by reformulating the non
local currents $J_i^\pm$ in terms of the physical fields; a quick
identification can be made using the dimensions $d=1+{1\over g_\rho}$
and $d=1+{1\over g_\sigma}$ (recall that for the separate  leads
fixed point, the approaching operators have $d={1\over
g_\rho}+{1\over g_\sigma}$, corresponding to the transfer of a
physical electron carrying a spin $1/2$, and $d={4\over g_\rho}$, 
corresponding to the transfer of a pair of electrons with opposite
spin).
Understanding  the physical meaning of these properties in more
details is an open challenge.  A possibility is that particles with
half the electron charge and half its spin are actually transfered: ``half
electrons'' have actually been encountered in the double barrier 
case, where their appearance corresponds to the transfer through 
an ``island''. To make further progress, a reliable formula for the boundary 
entropy would
be most useful - technical difficulties have to be resolved 
before this  can be obtained, however. 

It is not clear to us how the existence of these new fixed points
changes the
conclusions of \cite{kanefish,AW}. Our analysis, however, indicates
that
an instanton approach to the study of the problem might be
misleading. In particular, the duality property, which is the main
ingredient of the
 lore  in the spinless case (although only partially understood yet),
cannot always hold  in the case with spin: the UV and IR expansions
of our  current do not match under the transformations $g\to 4/g$. 

\bigskip
\noindent{\bf Acknowledgments} We thank I. Affleck, P. Fendley, M. P. Fisher
 and N. Warner  for
useful discussions. This work was upported by the Packard Foundation, the
 National Young
Investigator Program and the DOE. It was also partly 
 supported by the NSF under Grant No. PHY94-07194. 

\appendix
\section{Non-integrability in the classical case.}

In this appendix we want to show that the system
described by the action
\begin{equation}
S=\int \sum_{i=1}^2 \frac{1}{2} (\partial_\mu\phi_i)^2 +
\Lambda \cos (\beta_1\phi_1)\cos (\beta_2\phi_2)
\end{equation}
does not have a spin 4  conserved quantity classically, and is presumably
non integrable. 

Using the light-cone
variables $z=x+iy$, $\bar{z}=x-iy$, the classical
equations of motion  are given by
\begin{eqnarray}
\label{classeq}
\partial_z\partial_{\bar{z}}\phi_1&=&-\frac{\Lambda \beta_1}{4}
\sin (\beta_1\phi_1) \cos (\beta_2\phi_2),  \nonumber \\
\partial_z\partial_{\bar{z}}\phi_2&=&-\frac{\Lambda \beta_2}{4}
\cos (\beta_1\phi_1) \sin (\beta_2\phi_2).
\end{eqnarray}
The first non trivial condition of integrabilty is the existence of
a spin 4 conserved current, $T_4$ such
that
\begin{equation}
\label{classint}
\partial_{\bar{z}}T_4=\partial_z \theta_2 .
\end{equation}
The most general form of the $T_4$ current respecting the
$U(1)$ symmetries is given by
\begin{equation}
T_4=A_1 (\partial_z\phi_1)^4+A_2 (\partial_z\phi_2)^4+
A_3 (\partial_z\phi_1)^2 (\partial_z\phi_2)^2+A_4
(\partial_z^2\phi_1)^2
+A_5 (\partial_z^2\phi_2)^2.
\end{equation}
When we take the $\partial_{\bar{z}}$ derivative of $T_4$, several
of the  terms can be eliminated up to total $\partial_z$ derivatives,
so it is enough to look for  a $\theta_2$  of the  form
\begin{eqnarray}
\theta_2=B_1 (\partial_z\phi_1)^2 &&\cos(\beta_1\phi_1)
\cos(\beta_2\phi_2)+
B_2 (\partial_z\phi_2)^2 \cos(\beta_1\phi_1) \cos(\beta_2\phi_2)+
 \nonumber \\
&& B_3(\partial_z\phi_1)(\partial_z\phi_2)\sin (\beta_1\phi_1)
\sin(\beta_2 \phi_2).
\end{eqnarray}
Straightforward
algebra gives the relations
\begin{eqnarray}
-A_1 \Lambda \beta_1=-\beta_1 B_1, & &\ \
-A_2 \Lambda\beta_2=-\beta_2 B_2 \nonumber \\
-A_3 \frac{\Lambda \beta_1}{2}=-\beta_1 B_2+\beta_2 B_3, & &\ \
-A_3\frac{\Lambda\beta_2}{2}=-\beta_2 B_1+\beta_1 B_3 \\
-A_4 \frac{\Lambda \beta_1^2}{2}=2 B_1, & &\ \
A_4 \frac{\Lambda \beta_1\beta_2}{2}=B_3 \nonumber \\
A_5 \frac{\Lambda\beta_1\beta_2}{2}=B_3, & & \ \
-A_5 \frac{\Lambda \beta_2^2}{2}=2 B_2. \nonumber
\end{eqnarray}
If $\beta_1\neq \beta_2$, then
these equations have no solutions and thus the system has no
non trivial conserved quantity of spin 4 classically. Presumably, it
then does not have conserved quantities of higher spin either.

The algebra in the quantum case is a bit different. The problem is to
find a chiral quantity   $T_4$, of the most general form
\begin{equation}
T_4=A_1 (\partial_z\varphi_1)^4+A_2 (\partial_z\varphi_2)^4+
A_3 (\partial_z\varphi_1)^2 (\partial_z\varphi_2)^2+A_4
(\partial_z^2\varphi_1)^2
+A_5 (\partial_z^2\varphi_2)^2.
\end{equation}
such that the residue of the simple pole
in the  short distance expansion  with the perturbation
is a total derivative. Using the propagators
$<\varphi_i(z)\varphi_j(z')>={-\delta_{ij}\over 4\pi}\ln (z-z')$,
one finds that  the quantity $T_4$   as given in the text
(\ref{tefour})
solves the problem.

That some $1+1$ quantum field theories can be non integrable classically, but 
are quantum mechanically integrable at finite values of  Planck's constant 
($\beta_i$ here)
was observed before: see for instance \cite{shankar}. Usually however,
quantum integrability becomes ``classical'' once the theory is fermionized. 
While a
similar phenomenon occurs here for the supersymmetric case
\cite{kobayashi}, we are not aware of
a classically integrable, partially fermionized, version of our model
for $\beta_1^2+\beta_2^2=4\pi$.

\section{Boundary quantum integrability}

We show that the boundary perturbation in the action (\ref{action})
preserves quantum integrability.
It is sufficient to prove that the boundary perturbation of its
conformal limit preserves the bulk integrable manifolds.
The proof follows Ghoshal and A.Z. Zamolodchikov argument in
\cite{GZ}.
The unperturbed conformal theory is constituted by the two
independent
bosons $\phi_1,\phi_2$  in the half plane, with  Neumann boundary
conditions
\begin{equation}
\frac{\partial}{\partial x} \phi_i (0,y) = 0
\end{equation}
such that the propagators turn out to be
\begin{equation}
\langle \phi_i (x_1,y_1) \phi_{j} (x_2,y_2) \rangle =
- \delta_{ij} \left[
\frac{1}{4\pi} \log\frac{(y_1-y_2)^2 + (x_1-x_2)^2}{\kappa^2}
+ \frac{1}{4\pi} \log\frac{(y_1-y_2)^2 +
(x_1+x_2)^2}{\kappa^2}\right],
\end{equation}
where $\kappa$ is a short distance cutoff. The half-line bosons
$\phi_i (x,y)$ can thus be considered
as the folding of two full line bosons $\Phi_i (x,y)$,
\begin{equation}
\phi_i (x,y) = \frac{1}{\sqrt{2}}
\left[\Phi_i (x,y) + \Phi_i (-x,y) \right].
\end{equation}
The full line  left and right components
$\Phi_i (x,y) = \Phi_{i,L} (x,y) + \Phi_{i,R} (x,y) $
allow the further decomposition of the half-line bosons in a left and
a right
component
\begin{equation}
\phi_i (x,y) = \Phi^e_i (x,y) + \Phi^e_i (-x,y)
\end{equation}
where we define the even fields
\begin{equation}
\Phi^e_i (x,y) = \frac{1}{\sqrt{2}}
\left[ \Phi_{i,L}(x,y) + \Phi_{i,R}(-x,y) \right]
\end{equation}
whose correlations are
\begin{equation}
\langle \Phi^e_i (x_1,y_1) \Phi^e_{j} (x_2,y_2) \rangle =
- \delta_{ij} \frac{1}{4\pi} \log\frac{y_1-y_2 + i
(x_1-x_2)}{\kappa}.
\end{equation}

Conformal invariance of the half-line model is guaranteed by the
condition
that no energy-momentum flux escapes the system across the boundary,
$T(x,y) = \overline{T}(-x,y)$, where
$T(x,y) = - 2\pi \sum_i \left(\partial_z \phi_i \right)^2 =
- 2\pi \sum_i \left(\partial_z \Phi^e_i (x,y) \right)^2$,
$\overline{T}(x,y) = - 2\pi \sum_i \left(\partial_{\bar{z}} \phi_i
\right)^2 =
- 2\pi \sum_i \left(\partial_{\bar{z}} \Phi^e_i (-x,y) \right)^2$,
 and $z=y + i x$.

This condition implies an infinite number of higher spin conservation
laws,
$T_{s+1} (x,y) = \overline{T}_{s+1} (-x,y)$, where the spin label $s$
is an odd integer. Therefore a boundary perturbation preserves
integrability if some of the conserved currents satisfy
\begin{equation}
\left.T_{s+1} (x,y) - \overline{T}_{s+1} (-x,y)\right|_{x=0} =
\frac{\partial}{\partial x} t_s(x).
\end{equation}
The standard  argument \cite{GZ} consists in evaluating the
left-hand side at first order in conformal perturbation theory
\begin{equation}
\lambda \int dy' \left[T_{s+1} (x,y) - \bar{T}_{s+1} (-x,y)\right]
\Phi(0,y'),
\label{blhs}
\end{equation}
where $\Phi(0,y) = \cos\beta_1/2\phi_1(y) \cos \beta_2/2\phi_2(y)$ is the
perturbing
operator.
Therefore the OPE of the currents with the perturbing operator are
needed. Since the half-line bosons at the boundary satisfy
\begin{equation}
\phi_i (0,y) = 2 \Phi_i (0,y)
\end{equation}
it turns out that the computation of (\ref{blhs}) is identical to the
one in the
bulk for the quantity  $\partial_{\bar{z}} T_{s+1}$ and the
perturbing operator
$\cos\beta_1\phi_1 (0,y)
\cos\beta_2\phi_2(0,y)$.
As a result, the bulk integrable manifold gives rise to a boundary
integrable one.

\section{Perturbative computation of the S matrix}

In this appendix we give more credibility to the S matrix
proposed in the text by checking that it is correctly
recovered in perturbation.
As shown in \cite{Fateev},
the bosonic action of (\ref{bulkact}) can be put in a useful fermionic 
form by first introducing two auxiliary boson fields 
$\chi_{{}_{1,2}} = \frac{1}{2\sqrt{\pi}} 
(\beta_1\phi_1 \pm \beta_2\phi_2)$ whose action  
\begin{equation}
S=\int\left\{\left(\frac{\pi}{2\beta_1^2} + \frac{\pi}{2\beta_2^2}\right)
[(\partial\chi_1)^2 + (\partial\chi_2)^2] + 
\left(\frac{\pi}{2\beta_1^2}-\frac{\pi}{2\beta_1^2}\right) 
(\partial\chi_1) (\partial\chi_2) + \Lambda \left(\cos 2\sqrt{\pi}\chi_1 +
\cos 2\sqrt{\pi}\chi_2\right)\right\}
\end{equation}
is the bosonization of the two fermions theory defined by
\begin{equation}
S=\int\left\{\sum_{i=1,2} \bar{\psi_i}
(i\gamma_{\mu}\partial^{\mu} - m)\psi_i +
g\sum_{i=1,2} (\bar{\psi_i}\gamma_{\mu}\psi_i)^2 + 
g_1 (\bar{\psi_1}\gamma_{\mu}\psi_1)
(\bar{\psi_2}\gamma^{\mu}\psi_2)\right\}.
\end{equation}
The couplings are defined by $g=\frac{\pi}{2}
(\frac{\pi}{\beta_1^2}+\frac{\pi}{\beta_2^2} -1)$ and 
$g_1=\pi (\frac{\pi}{\beta_1^2} - \frac{\pi}{\beta_2^2})$.
The integrable manifold (\ref{cond}) becomes 
\begin{equation}
g_1^2 = 2 g (2 g + \pi).
\label{manifoldg}
\end{equation}

The perturbative computation of the scattering amplitudes of the fermions
permits their identification with the composite charge and spin solitons.
Denoting by $(\rho \,\,\sigma)$ with $\rho,\sigma =\pm$ the four 
elementary excitations whose exact scattering matrix
is (\ref{SSmatrix}), the identification is 
\begin{equation}
\psi_1 = (+ +)\,,\,\,\,\bar{\psi_1}= (- -)\,,\,\,\,
\psi_2 = (+ -)\,,\,\,\,\bar{\psi_2}= (- +).
\end{equation}
The perturbative region $g\sim 0$ for the two fermions theory 
implies $\beta_i^2 \sim 2\pi$ and therefore each sine-Gordon 
component in the exact solution (\ref{SSmatrix}) 
\begin{equation}
S_{(\rho_1\sigma_1) (\rho_2\sigma_2)}^{(\rho'_1\sigma'_1) 
(\rho'_2\sigma'_2)} (\theta) = - S_{\rho_1\rho_2}^{\rho'_1 \rho'_2}
(\theta;\hat{\beta_1}) 
S_{\sigma_1 \sigma_2}^{\sigma'_1 \sigma'_2}(\theta;\hat{\beta_2})
\end{equation}
can be expanded \cite{Weisz}
around the respective free fermion points $\hat{\beta_i}^2 \sim 4\pi$.
We recall that the sine-Gordon couplings $\hat{\beta_i}$ 
are related to the couplings $\beta_i$ of our model by 
$\hat{\beta_i}^2=\frac{8\pi\beta_i^2}{\beta_i^2 + 2\pi}$. 
As an example let us consider the following expansion
\begin{equation}
S_{(+ +) (- -)}^{(- -) (+ +)}(\theta)= \pi^2 (\frac{2\pi}{\beta_1^2} -1) 
(\frac{2\pi}{\beta_2^2} - 1) \frac{1}{\sinh^2\theta} + o(2)
= -\frac{2\pi g}{\sinh^2\theta} + o(2)
\label{expansion}
\end{equation}
where we have also used the integrability condition (\ref{manifoldg}).
This result has to be compared with the direct perturbative evaluation of the
amplitude $S_{\psi_1,\bar{\psi_1}}^{\bar{\psi_1} \psi_2}$.
We will proceed with a dispersion relation method, 
as described in \cite{Weisz} for the Thirring model. 
The two perturbative couplings 
$g$ and $g_1$ are treated as independent. 
The integrability condition will be imposed at the end of the 
second order computation which is needed because
$o(g_1^2)$ still contribute
to the lowest order in $g$ once (\ref{manifoldg}) is taken into account.
The tree level amplitudes, schematically in figure 1, are determined by
the two four fermion vertices in the lagrangian contracted on the respective
external fermions. 
\begin{figure}
\vglue 0.4cm\epsfxsize 4cm\centerline{\epsfbox{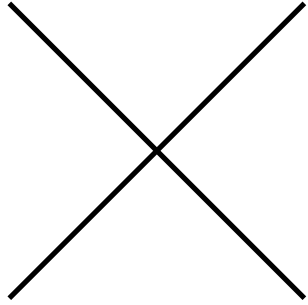}}\vglue 0.4cm
\caption[]{\label{fig1} Tree level diagram.}
\end{figure}
Some of them are 
\begin{eqnarray}
S_{\psi_1 \psi_1}^{\psi_1 \psi_1}(\theta) &=& - 
(1 + i 2 g \frac{1-\cosh\theta}{\sinh\theta}) \hspace{2cm}
 S_{\psi_1 \bar{\psi_1}}^{\psi_1 \bar{\psi_1}}(\theta) = - (1 + i 2 g 
\frac{1+\cosh\theta}{\sinh\theta}) 
\\ \nonumber
S_{\psi_1 \bar{\psi_1}}^{\bar{\psi_1} \psi_1}(\theta) &=&
 i4g\frac{1}{\sinh\theta} 
\\ \nonumber 
 S_{\psi_1 \bar{\psi_1}}^{\psi_2 \bar{\psi_2}}(\theta) &=&  
- i  g_1 \frac{1}{\sinh\theta} \hspace{3.55cm}
 S_{\psi_1 \psi_2}^{\psi_2 \psi_1}(\theta) = - i  g_1 \frac{1}{\sinh\theta}  
\\ \nonumber 
S_{\psi_1 \bar{\psi_1}}^{\bar{\psi_2} \psi_2}(\theta) &=&
  i  g_1 \frac{1}{\sinh\theta}  \hspace{3.8cm}
S_{\psi_1 \bar{\psi_2}}^{\bar{\psi_2} \psi_1}(\theta) = 
 i  g_1 \frac{1}{\sinh\theta} 
\\ \nonumber 
S_{\psi_1 \psi_2}^{\psi_1 \psi_2}(\theta) &=& 
- (1 - i  g_1 \frac{\cosh\theta}{\sinh\theta}) \hspace{2.6cm}
 S_{\psi_1 \bar{\psi_2}}^{\psi_1 \bar{\psi_2}}(\theta) = - (1 + i  g_1 
\frac{\cosh\theta}{\sinh\theta}). 
\end{eqnarray}
We have written on the same lines the crossing symmetric pairs.
The other amplitudes can be obtained by the listed above exploiting
either the symmetry $1 \leftrightarrow 2$ or CPT symmetry.

The real part of the second order contributions, schematically 
in figure 2, can be obtained by the cutting technique. 
\begin{figure}
\vglue 0.4cm\epsfxsize 4cm\centerline{\epsfbox{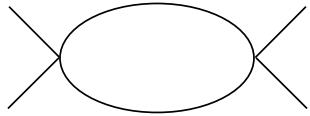}}\vglue 0.4cm
\caption[]{\label{fig2} Second order diagram.}
\end{figure}
Actually it is not necessary
to use explicitely the Cutkowski rule. 
One can use unitarity, as we show in the following example.
Let's consider the amplitude $S_{\psi_1 \bar{\psi_1}}^{\bar{\psi_1} \psi_1}$.
According to unitarity of the total S-matrix 
$S^\dagger S = 1$ and elasticity of $2 \rightarrow 2$ particles
scattering the real part 
of the second order contribution is determined by 
the tree level amplitudes
\begin{eqnarray}
\mbox{Re} (S_{\psi_1 \bar{\psi_1}}^{\bar{\psi_1} \psi_1}) &=&
\frac{1}{2} \left[ (S_{\psi_1 \bar{\psi_1}}^{\psi_1 \bar{\psi_1}} + 1)
(S_{\psi_1 \bar{\psi_1}}^{\bar{\psi_1} \psi_1})^* + 
S_{\psi_1 \bar{\psi_1}}^{\bar{\psi_1} \psi_1} 
(S_{\bar{\psi_1} \psi_1}^{\bar{\psi_1} \psi_1} +1)^* +
S_{\psi_1 \bar{\psi_1}}^{\psi_2 \bar{\psi_2}} 
(S_{\psi_2 \bar{\psi_2}}^{\bar{\psi_1} \psi_1})^* + 
S_{\psi_1 \bar{\psi_1}}^{\bar{\psi_2} \psi_2} 
(S_{\bar{\psi_2} \psi_2}^{\bar{\psi_1} \psi_1})^* \right] + o(3)
\\ \nonumber
&=&
 - 8 g^2 \frac{1+\cosh\theta}{\sinh^2\theta}  - g_1^2 \frac{1}{\sinh^2\theta} 
+o(3) .
\label{realpart}
\end{eqnarray}
In total generality a second order diagram in the class of figure 2
can be parameterized as
\begin{equation}
g^2 \,\,\,\frac{i}{\sinh\theta} 
\,\,\,\frac{i\pi-\hat\theta}{\pi\sinh\hat\theta}\,\,\,
[a\cosh^2\theta +b\cosh\theta +c]
\end{equation}
where $\hat\theta=\theta,\,\,i\pi-\theta,\,\,i\pi$ 
when the total momentum $P$ which
flows into the diagram is $P=(p_1+p_2),\,\,\pm(p_1-p_2),\,\,0$, resp.
Obviously the order $g^2$ can be substituted by 
$g_1^2$ and $g g_1$ depending on the kind of vertices
 involved in the diagram.
The factor $1/{\sinh\theta}$ is the usual jacobian of the total 
momentum
conservation expressed in the rapidities $\theta_1,\theta_2$ 
and the mandelstam
variable is $s=(p_1+p_2)^2=2m^2 (1+\cosh\theta)$.
The function $r(\hat\theta)=\frac{i\pi-\hat\theta}{\pi\sinh\hat\theta}$
is the result of the dispersion integral on the two-particles 
phase space $\Theta(s - 4m^2)/\sqrt{s(s-4m^2)}$ 
\begin{equation}
r(\hat\theta)=\frac{1}{\pi}\int_{4 m^2}^{\infty}\frac{ds'}{s'- P^2}
\frac{2 m^2}{\sqrt{s'(s'- 4 m^2)}}. 
\end{equation}
The polynomial in $\cosh\theta$ is the effect of the contraction
 of external fermion wave functions on the numerators of the loop
 fermion propagators.
Each diagram can at most be logarithmic divergent at high 
energy, $\log s \sim\theta$.
This fact constraints the polynomial to be of order not greater than two.
Therefore the amplitude of our example can be written in the form
\begin{eqnarray}
S_{\psi_1 \bar{\psi_1}}^{\bar{\psi_1}\psi_1}(\theta) &=&
\frac{i 4 g}{\sinh\theta} +\\\nonumber
&&\frac{i 8 g^2}{\sinh\theta} 
\left\{r(\theta) [A\cosh^2\theta + B\cosh\theta + C] + 
r(i\pi-\theta) [\bar{A}\cosh^2\theta + \bar{B}\cosh\theta + \bar{C}] +
r(i\pi) [D\cosh\theta +E]\right\}+\\\nonumber
&&\frac{i g_1^2}{\sinh\theta}
\left\{r(\theta) [A_1\cosh^2\theta + B_1\cosh\theta + C_1] + 
r(i\pi-\theta) [\bar{A}_1\cosh^2\theta + \bar{B}_1\cosh\theta + \bar{C}_1] +
r(i\pi) [D_1\cosh\theta +E_1]\right\} \\\nonumber
&& + o(3).
\end{eqnarray}
Some of the coefficients of the polynomials, which encode the 
freedom in the finite 
part of a regularized diagram, are determined by imposing i) the matching
with the already computed real part (\ref{realpart}), ii) the
 asymptotic behaviour 
$S_{\psi_1 \bar{\psi_1}}^{\bar{\psi_1}\psi_1}(\theta)\rightarrow 0$
 as $\theta\rightarrow \infty$ and 
iii) the crossing symmetry relation 
$S_{\psi_1 \bar{\psi_1}}^{\bar{\psi_1}\psi_1}(\theta)=
S_{\psi_1 \bar{\psi_1}}^{\bar{\psi_1}\psi_1}(i\pi-\theta)$.
The result is 
\begin{equation}
S_{\psi_1 \bar{\psi_1}}^{\bar{\psi_1}\psi_1}(\theta)=
\frac{i 4 g}{\sinh\theta} -\frac{ 8 g^2}{\sinh^2\theta}(1+\cosh\theta)
-\frac{ g_1^2}{\sinh^2\theta} -
\frac{i 8 g^2}{\pi\sinh\theta} \left(2 \frac{\theta}{\sinh\theta}
 \cosh\theta -E\right)
+ \frac{i g_1^2}{\pi\sinh\theta} E_1  + o(3).
\end{equation}

The same procedure can be used for the amplitudes 
$S_{\psi_1 \psi_1}^{\psi_1\psi_1}$ and
$S_{\psi_1 \bar{\psi_1}}^{\psi_1 \bar{\psi_1}}$ which are 
mapped one into the other by crossing symmetry and whose 
asymptotic behaviour at high energy is
$S(\theta) \rightarrow \makebox{constant}$. 
We obtain
\begin{eqnarray}
S_{\psi_1 \psi_1}^{\psi_1\psi_1}(\theta) &=&
-1-\frac{i 2 g}{\sinh\theta} (1-\cosh\theta) +
\frac{2g^2}{\sinh^2\theta}(1-\cosh\theta)^2 \\\nonumber
&&-\frac{i 2 g^2}{\pi\sinh\theta} \left(4\frac{\theta}{\sinh\theta} 
+d\cosh\theta +e\right)
-\frac{i g_1^2}{\pi\sinh\theta} \left(\frac{\theta}{\sinh\theta}
+d_1\cosh\theta +e_1\right) + o(3)\\\nonumber
S_{\psi_1 \bar{\psi_1}}^{\psi_1\bar{\psi_1}}(\theta) &=&
-1-\frac{i 2 g}{\sinh\theta} (1+\cosh\theta) 
+\frac{2g^2}{\sinh^2\theta}[(1+\cosh\theta)^2 +4]
+\frac{g_1^2}{\sinh^2\theta}\\\nonumber 
&&-\frac{i 2 g^2}{\pi\sinh\theta} \left(-4\frac{\theta}{\sinh\theta}
 -d\cosh\theta +e\right)
-\frac{i g_1^2}{\pi\sinh\theta} \left(-\frac{\theta}{\sinh\theta}
-d_1\cosh\theta +e_1\right) + o(3).
\end{eqnarray}

Imposing now the threshold behaviour 
$\lim_{\theta\rightarrow 0}S_{\psi_1 \psi_1}^{\psi_1\psi_1}(\theta)=-1$ 
and
$\lim_{\theta\rightarrow 0}
(S_{\psi_1 \bar{\psi_1}}^{\psi_1\bar{\psi_1}}(\theta) +
S_{\psi_1 \bar{\psi_1}}^{\bar{\psi_1}\psi_1}(\theta)) =0$
we obtain the four relations
\begin{equation}
E=-\frac{d}{2},\,\,\,\,\,\,\,\,e=-4-d,\,\,\,\,\,\,\,\,E_1 = -2 -2 d_1,
\,\,\,\,\,\,\,\,e_1=-1-d_1.
\end{equation}
We are thus left with only two undetermined constants, $d$ and $d_1$.
This freedom reflects the freedom we have in the definition 
of the coupling constants $g$ and $g_1$. Let us choose to define them 
through the asymptotic behaviour of the amplitudes
\begin{eqnarray}
\lim_{\theta\rightarrow +\infty} 
S_{\psi_1 \psi_1}^{\psi_1\psi_1}(\theta) &=&
- e^{-i 2 g}\\\nonumber
\lim_{\theta\rightarrow +\infty} 
S_{\psi_1 \psi_2}^{\psi_1 \psi_2}(\theta) &=&
- e^{-i g_1}
\end{eqnarray} 
fixing the coefficients to $d=0$, $d_1=0$.

The amplitude we want to compare with the expansion of the exact results 
(\ref{SSmatrix})
is therefore
\begin{equation}
S_{\psi_1 \bar{\psi_1}}^{\bar{\psi_1}\psi_1}(\theta)=
\frac{i 4 g}{\sinh\theta} -\frac{ 8 g^2}{\sinh^2\theta}(1+\cosh\theta)
-\frac{ g_1^2}{\sinh^2\theta} - 
\frac{i 16 g^2}{\pi\sinh\theta} \frac{\theta}{\sinh\theta} \cosh\theta
- \frac{i 2 g_1^2}{\pi\sinh\theta}+o(3).
\end{equation}
On the integrable manifold (\ref{manifoldg}) at lowest order in $g$ we 
obtain
\begin{equation}
S_{\psi_1 \bar{\psi_1}}^{\bar{\psi_1}\psi_1}(\theta)=
-\frac{2\pi g}{\sinh^2\theta} + o(2)
\end{equation}
which coincides with the lowest order expansion of the exact result 
(\ref{expansion}).
The same check can now be extended to the remaining amplitudes.

\section{Keldysh Computation.}

In order to compute the differential conductance we need to use
the Keldysh formalism since the system is driven by reservoirs.
To do so, we will use the formulation of the model on the full line
as described by the equations (\ref{lateruse}).  The effect of the
reservoirs can be implemented by shifting the charge fields,
$\Theta_\rho\rightarrow \Theta_\rho+\sqrt{g_\rho/\pi} a(t)/2$ with
$\partial_t a(t)=V$.  Under this prescription the current is
evaluated by taking the functionnal derivative
\begin{equation}
I(t)=-i \frac{\delta \log Z[a(t)]}{\delta a(t)},
\end{equation}
where we have used conventions in which $\hbar=e=1$.
Using the Keldysh contour, ${\cal C}$, which goes from
$-\infty$ to $\infty$ and then comes back (see figure 3), 
to expand the
partition function we obtain to first order
\begin{figure}
\vglue 0.4cm\epsfxsize 6cm\centerline{\epsfbox{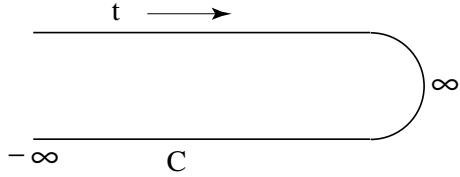}}\vglue 0.4cm
\caption[]{\label{fig4} Contour for the perturbative evaluation.}
\end{figure}
\begin{equation}
I^{(2)}(0)=-i \frac{\lambda^2 g_\rho}{8}\int_{\cal C}dt  \
\sin[\frac{g_\rho}{2}a(t)] \left\{
P_{g_\rho+g_\sigma}^{\mu+}(t)+P_{g_\rho+g_\sigma}^{+\mu}(-t)\right\}
\end{equation}
where $\mu=\pm$ depending on the location of $t$, ie upper part of
the
contour or lower part of the contour.  The functions
$P^{\mu,\mu'}_g(t)$
are the corresponding contraction of the vertex operators time
ordered on the Keldysh contour
\begin{equation}
P^{\pm\pm}_g(t)={1\over (\delta\pm \vert t\vert)^{g/2}}, \ \
P^{\pm \mp}_g(t)={1\over (\delta \mp t)^{g/2}}.
\end{equation}
To this order the result can be found explicitely, it is given by
\begin{equation}
I^{(2)}(0)=-\frac{\pi}{2 \Gamma[(g_\rho+g_\sigma)/2]}
\left(\frac{g_\rho}{2}\right)^{(g_\rho+g_\sigma)/2}
\vert \lambda \vert^2 V^{\frac{g_\rho+g_\sigma}{2}-1} .
\end{equation}
This is in agreement with the Bethe ansatz solution given in the
bulk of the text provided we make the identification
(putting $h=2\pi$ and $e=1$ in the TBA expressions and
$g_\rho+g_\sigma=2$ in the previous perturbative expression)
\begin{equation}
\vert \lambda \vert^2=\frac{2}{\pi} \sin(\frac{\pi}{2}g_\rho) T_B.
\end{equation}

\section{Operator content of the IR fixed point.}

To get a better knowledge of the operator content of the
IR fixed point, it is useful to study
correlation functions of various operators. This is done using
form-factors \cite{LSS}.  The form factors for  theories with a
factorized S matrix of the form (\ref{SSmatrix}) were analysed
by Smirnov \cite{smirss}
for the massive theory described in the bulk of the text.   These
are now matrix elements of operators ${\cal O}$ in the space
of states created by the Fadeev-Zamolodchikov operators
\begin{equation}
\langle 0|{\cal O}(0,0) Z^*_{\epsilon_1}(\theta_1)
 \cdots Z^*_{\epsilon_n}(\theta_n)|0\rangle
\end{equation}
where now the asymptotic states are described by isotopic indices
giving the two quantum numbers of the excitation: $\epsilon_i=
(T_1,T_2)$.
Here we
will concentrate on the current form factors.  For these operators,
the massless limit, which is needed to treat the boundary
perturbation  problem, has basically the same form
has for the boundary sine-Gordon  case discussed  in \cite{LSS}.
Here, the theory has $U(1)\otimes U(1)$
symmetry, and we can
 construct currents invariant with respect to either
of these $U(1)$.
There is an infinity of non-zero matrix elements, or form factors
but for the sake of the argument we need only the two solitons
ones.  Let us concentrate on
the right moving currents. For  the  field $\phi_1$ one has
\begin{equation}
f_+^1(\theta,\theta')_{\epsilon,\epsilon'}=
d_{\beta_1,\beta_2} e^{(\theta+\theta')/2} { \zeta_{\beta_1,\beta_2}
(\theta-\theta')\over \cosh\frac{1}{2\beta_1}(\theta'-\theta-i\pi)
\sinh\frac{1}{2\beta_2}(\theta'-\theta-i \pi)} T_1
\end{equation}
and for the field $\phi_2$,
\begin{equation}
f_+^2(\theta,\theta')_{\epsilon,\epsilon'}=
d_{\beta_1,\beta_2} e^{(\theta+\theta')/2} { \zeta_{\beta_1,\beta_2}
(\theta-\theta')\over \sinh\frac{1}{2\beta_1}(\theta'-\theta-i\pi)
\cosh\frac{1}{\beta_2}(\theta'-\theta-i \pi)} T_2 .
\end{equation}
The constant $d_{\beta_1,\beta_2}$ and the function
$\zeta_{\beta_1,\beta_2}$ are given in \cite{smirss} and will
not be needed in the following.
The computation of current-current correlator in the Luttinger
model follows immediatly using the methods of \cite{LSS}.
By using the explicit expression written before, it is easy to
check that $<0|\partial_\pm\phi_1 \partial_\pm\phi_2|0>=0$
as expected.  The zero temperature, frequency dependent,
conductance would then follow from a correlator of the
form $<0|\partial_+\phi_1 \partial_-\phi_1|0>$,
after the usual identification $\theta_\rho\to\phi_1$.  Using the
boundary
state formulation, the previous form factors lead to a contribution
to that correlator which is of the form
\begin{eqnarray}
\int {d\theta d\theta'\over (2\pi)^2 2!} \vert
f^1_+(\theta,\theta')\vert^2
e^{(z-\bar{z})(e^{\theta}+e^{\theta'})} && \\ \nonumber
 [P_{\hat{\beta}_1}(\theta_{1B})
P_{\hat{\beta}_1}(\theta_{2B})&&-Q_{\hat{\beta}_1}(\theta_{1B})
Q_{\hat{\beta}_1}(\theta_{2B})]
[P_{\hat{\beta}_2}(\theta_{1B})P_{\hat{\beta}_2}
(\theta_{2B})+Q_{\hat{\beta}_2}(\theta_{1B})
Q_{\hat{\beta}_2}(\theta_{2B})],
\end{eqnarray}
where we used the notation $\theta_{iB}=\theta_i-\theta_B$ and
$\theta_B=\log T_B$.  $P_{\hat{\beta}}$ and $Q_{\hat{\beta}}$ are the
sine-Gordon
elements of the reflection matrix $R$ are coupling $g$.
We have $P=R^+_+=R^-_-$ and $Q=R_-^+=R_+^-$.
The frequency dependence follow from
a Fourier transformation of the previous expression
over time.  We then get one
single integral and the frenquency dependence can be recast
in the reflection matrices by shifting the rapidities as was done
in \cite{LSS}.  The expansion around the IR fixed point then
follows from the expansion of the reflection matrices.  Doing
this we obtain a multiple power series with the following
exponents: $(\omega/T_B)^{2n}$, $(\omega/T_B)^{4\pi n/\beta_1^2}$
and $(\omega/T_B)^{4\pi n/\beta_2^2}$.  Assuming the current
correlator explores the whole operator content of the theory, this is
consistent with (\ref{conji})

\end{document}